# Binary Formation in Planetesimal Disks
# I. Equal Mass Planetesimals

*Junko K. Daisaka

E-mail: kominami@cfca.jp

Division of Theoretical Astronomy,National Astronomical Observatory of Japan

Junichiro Makino

E-mail: makino@cfca.jp

Division of Theoretical Astronomy,National Astronomical Observatory of Japan

Hiroshi Daisaka

E-mail: hiroshi.daisaka@srv.cc.hit-u.ac.jp

Faculty of Commerce and Management,Hitotsubashi University

*former name Junko Kominami

Running Head: Binary Formation in Planetesimal Disks I.

Correspondent Address:

Junko K. Daisaka

Division of Theoretical Astronomy

National Astronomical Observatory of Japan

Osawa, 2-21-1, Mitaka, Tokyo 181-8588, Japan

tel: +81-422-34-3832

fax: +81-422-34-3746

e-mail: kominami@cfca.jp

# Abstract

As to April 2010, 48 TNO (trans-Neptunian Object) binaries have been found. This is about 6% of known TNOs. However, in previous theoretical studies of planetary formation in the TNO region, the effect of binary formation has been neglected. TNO binaries can be formed through a variety of mechanisms, such as three-body process, dynamical friction on two massive bodies, inelastic collisions between two bodies etc. Most of these mechanisms become more effective as the distance from the Sun increases. In this paper, we studied three-body process using direct $N$-body simulations. We systematically changed the distance from the Sun, the number density of planetesimals, and the radius of the planetesimals and studied the effect of the binaries on the collision rate of planetesimals. In the TNO region, binaries are involved in 1/3 - 1/2 of collisions, and the collision rate is increased by about a factor of a few compared to the theoretical estimate for the direct two-body collisions. Thus, it is possible that the binaries formed through three-body process significantly enhance the collision rate and reduce the growth time scale. In the terrestrial planet region, binaries are less important, because the ratio between the Hill radius and physical size of the planetesimals is relatively small. Although the time scale of our simulations is short, they clearly demonstrated that the accretion process in the TNO region is quite different from that in the terrestrial planet region. Simulations which cover longer time scale are required to obtain more accurate estimate for the accretion enhancement.

# 1 Introduction

Several mysteries of TNOs have emerged through the advance of observation. One of the mysteries is the formation mechanism of Pluto-sized object at ∼500 AU (trans-Neptunian object called Sedna). The time scale of in-situ formation of such a large body is longer than the age of Solar system. There should be some mechanism other than in-situ formation for such an object. Another mystery is the existence of eccentric, similar-sized planetesimal binaries. The separation of many TNO binaries is large (several hundred physical radius of one of the components). These characteristics of TNO binaries are quite different from those of main-belt asteroid binaries. Binaries in the main-belt asteroid have small separation and significantly different mass of components. The eccentricities of binaries in main-belt asteroids are small (∼ 0-0.1). Such binaries are thought to be formed through collisions of two planetesimals.

As described above, physical properties of binaries in the trans-Neptunian region are quite different from those of the binaries in main-belt asteroid. Hence, the formation mechanism of trans-Neptunian binaries(TNB) should be different from that of main-belt asteroid binaries. So far, five scenarios have been proposed. First, Weidenschilling (2002) proposed a mechanism that explains the formation of equal-mass binary. Within the Hill sphere of the primary, a collision of two other planetesimals takes place. The collided two bodies form a planetesimal with a size similar to the primary with low relative velocity. As a result, it forms a binary with the primary. Goldreich et al. (2002) proposed two mechanisms. Both mechanisms start with a temporal capture of a planetesimal within a Hill sphere of a primary. In one of the mechanisms, dynamical friction of the surrounding planetesimal sea takes away internal energy of the temporarily formed binary and makes it tighter. In the second mechanism, a third similar sized planetesimal takes away the internal energy by close-encounter (this mechanism is sometimes called three-body process). Forth mechanism was proposed by Funato et al. (2004). They explain the formation of highly-eccentric and similar-mass binaries. Two planetesimals collide and fragmentation takes place. Small fragment orbits around a large fragment, as main-belt asteroid binaries formed. Third planetesimal encounters the binary, exchanging the small fragment with itself. In Astakov et al. (2005), once a pair with planetesimals become loosely captured. The separation is about a Hill radius of one of the pair. The time scale of such a capture is long enough to allow other planetesimals to intervene. Interaction among these planetesimals causes a chaotic orbital evolution resulting in a binary formation with semimajor axis smaller than Hill radius of one of the planetesimals. There is no clear understanding concerning which of the proposed mechanism is the most important.

In order to understand the formation of TNBs and evolution of TNO region, the dynamic and accretion mechanism of planetesimals are needed to be studied. Carrying out global $N$-body simulation is one of the ways to understand the dynamics. One of the reasons that global $N$-body simulation of outer region of Solar system has been avoided is that it is too compute-intensive. Global $N$-body simulation needs large number of particles. In addition, the time scale of orbital evolution is long compared to that of the inner region.

Global $N$-body simulations of terrestrial planet region has been carried out and the planets' growth time scale has been analytically estimated (e.g. Kokubo and Ida 2002). Runaway growth and oligarchic growth of planetesimals (Kokubo and Ida 1998, 2000) result in Mars-sized protoplanet formation. Such protoplanets' orbits become unstable as majority of gas dissipates. The protoplanets coagulate and form terrestrial planets. If this scenario is applied to the outer region of Solar system, the in-situ formation time scale of Neptune would be longer than the age of Solar system ($\sim 10^9$ years) (Kokubo and Ida 2002). More over, time scale of in-situ formation of Sedna reaches $\sim 10^{10}$ years, if the same scenario is applied. Of course, it is unpractical to carry out such a long $N$-body simulation starting with $\sim 10^5$-$10^6$ particles.

$N$-body simulation is not the only way to explore the coagulation evolution of planetesimals. Kenyon and Luu (1998, 1999) used coagulation equation to follow the growth of 100 - 1000 km-sized planetesimals. They showed that several 100 km-sized planetesimals could be formed in $\sim 10^8$ years in TNO region, if the surface density of the disk is enhanced by a factor of 2-4 (Kenyon and Luu 1999). The coagulation equation they used did not include the effect of planetesimal binary formation.

The main goal of series of this paper is to study the dynamics and accretion mechanism in TNO region. In order to understand the mechanisms, first, we need to see how many planetesimal binaries are formed in TNO region through what mechanism. Secondly, the effect of binaries to the accretion process of planetesimals need to be understood. Although binaries might affect the growth of planetesimals, as in star clusters (Portegies Zwart et al. 1999), no systematic study of the effect of binary has been done so far.

In this paper, we report the result of global $N$-body simulation of planetesimal disk around the Sun. As the first step, we start with equal-mass planetesimals. The maximum initial number of planetesimals is $\sim 10^5$. Our work shows that binary formation through three-body process (the mechanism discussed in Goldreich et al. 2002) is fairly common. In addition, the accretion rate of planetesimals was enhanced by binary formation by factor of a few. In section 2, we show the analytical estimate of binary formation and the collision enhancement due to the binaries. The model and initial conditions are explained in section 3. Section 4 describes the result, and the last section summerize this paper.

# 2 Theoretical Estimate

In this section, we estimate the effect of of binaries on the collision rate. In section 2.1, we estimate the difference of collision cross section of a single planetesimal and a binary. In section 2.2, we estimate the formation rate binaries through three-body process. In section 2.3, by combining the results in sections 2.1 and 2.2, we estimate the enhancement of the collision rate due to the formation of binaries.

## 2.1 Estimate of Cross Section

For a single planetesimal, the cross section of collision ($\sigma_{\text{cross},1}$) is written as (Ida and Nakazawa 1989)

$$\sigma_{\text{cross},1} = \pi R_{\text{p}}^2 \left[1 + \left(\frac{v_{\text{esc}}}{v_{\text{rel}}}\right)^2\right], \tag{2.1}$$

where $R_{\text{p}}$ is physical radius of the planetesimal, $v_{\text{esc}}$ is the escape velocity and $v_{\text{rel}}$ is the relative velocity ($ev_{\text{kep}}$, $e$ is the eccentricity and $v_{\text{kep}}$ is the Kepler velocity). When a binary experiences a close encounter with a third body, three bodies would undergo complex resonant interaction. During such interaction, a collision between two bodies can occur. The cross section of such three-body collisions($\sigma_{\text{cross},2}$) is estimated by Hut and Inagaki (1985) as

$$\sigma_{\text{cross},2} = 8.5 \frac{\pi a_{\text{b}}^2}{v_{\text{rel}}^2} \left(\frac{3}{2}\frac{Gm}{a_{\text{b}}}\right) \left(\frac{R_{\text{p}}}{a_{\text{b}}}\right)^{0.4} \tag{2.2}$$

where $v_{\text{rel}}$ is the relative velocity between a single star and a binary before the encounter, $a_{\text{b}}$ is the semimajor axis of the binary and $m$ is the mass of a planetesimal [see Hut and Inagaki (1985) for detail]. Hence, the ratio of the cross sections is

$$\frac{\sigma_{\text{cross},2}}{\sigma_{\text{cross},1}} \sim 6.5 \left(\frac{a_{\text{b}}}{R_{\text{p}}}\right)^{0.6}. \tag{2.3}$$

This ratio is shown in fig.1, for $a_{\text{b}} = r_{\text{H}}$(solid line) and $a_{\text{b}} = 0.1 r_{\text{H}}$. The value $\sigma_{\text{cross},2}$ can be much larger than $\sigma_{\text{cross},1}$ in TNO region. Second, a binary, once formed, changes its eccentricity through distant encounters with other bodies, either through Kozai mechanism (Kozai 1962) or through random close encounters. When the eccentricity of a binary becomes high enough, its two components can collide with each other. In addition, when a binary is formed, the barycentric mass of the binary increases and $v_{enc}$ in $\sigma_{cross,1}$ increases as well. Meanwhile, since the mass of a binary is larger than that of a single planetesimal, its random velocity becomes smaller, resulting in smaller $v_{rel}$. Hence, $\sigma_{cross,1}$ increases when a binary is formed. If the cross section $\sigma_{cross,2}$ is added to the conventional theory (Weidenschilling 2002) it would increase the accretion rate.

Above argument is based on three-body process. As we overviewed in the introduction, other mechanisms have been proposed (e.g. Weidenschilling 2002, Goldreich et al. 2002, Funato et al. 2004). If the collision rate is enhanced by the three-body process, it might be enhanced by other mechanisms as well. It is possible that the growth time scale of planetesimals can be affected by all of the five mechanisms. Investigation of effects of other mechanisms will be carried out in future studies.

## 2.2 Estimate of Binary Formation Rate

Here we estimate the binary formation rate of three-body process. Hard binary formation rate in star clusters has already been investigated thoroughly in past literature (e.g. Goodman and Hut 1993). They found that the formation rate per unit volume $\dot{n}_{b,s}$ is given by

$$\dot{n}_{b,s} \sim G^5 n^3 m^5 \sigma^{-9} \tag{2.4}$$

where $G$ is the gravitational constant, $n$ is the number density of single stars, $m$ is the mass of a single star, and $\sigma$ is the velocity dispersion. This formula is based on the solution of the integral equation which expresses formation, destruction, hardening and softening of binaries. This formula gives the same dependence as that in the more recent estimate by SS08 for the specific model considered by them, and can be applicable to a wider range of conditions. Here we present a rough estimate of a planetesimal binary formation using eq.(2.4).

Assuming that the surface density scales as that of minimum mass Solar nebula (MMSN, $\Sigma \propto a^{-1.5}$, where $\Sigma$ is the surface density and $a$ is the semimajor axis), the scale height is proportional to $a^{1/2}$. Hence, the number density becomes proportional to $a^{-2}$. The velocity distribution is $\sigma \sim e v_{kep}$, where $e$ is the eccentricity and $v_{kep}$ is Kepler velocity, with $e$ being a constant. Therefore, the binary formation rate per unit volume is proportional to $a^{-3/2}$. In addition, if we assume a width of the ring $\Delta a$ is proportional to its inner semimajor axis $a_{\text{inner}}$ and inclination being constant, the volume of the ring is proportional to $a^3$. Hence, the binary formation rate for a ring with constant $\Delta a/a$ is proportional to $a^{3/2}$, for the outer region where the ratio between the Hill radius and the physical radius of the planetesimals is large enough to allow formation of three-body binaries. From eq.(2.5), the rate of the direct collision in the same region is proportional to $a^{-3/2}$. Thus the rate of binary formation relative to the direct collision increases as $a^{1.5}$.

We assumed that $e$ does not depend on the distance from the Sun, which is clearly an oversimplification since the velocity dispersion would depend on the ratio between the Hill radius and physical radius, through the equilibrium between the viscous stirring, collisional damping and gas drag. However, since we consider the effect of binaries to the collision rate itself, we did not use the equilibrium value of velocity dispersion obtained assuming that the binaries have no effect on the collision rate. In addition, it would take very long time to establish the equilibrium in the TNO region, though the time scale is fairly short at 1AU.

## 2.3 Collision Enhancement due to Binary Formation

From eq.(2.2), we can see that, once a binary is formed, the time scale of collision with a third star is quite short. Therefore, we can assume the rate of binary induced collisions is the same as the binary formation rate, which is proportional to $a^{3/2}$ for a ring with $\Delta a \propto a$. For direct collisions, the collision rate per unit volume is given by (Ida and Nakazawa 1989)

$$\dot{n}_{\text{coll}} \sim \rho^{-1/3} G n^2 m^{4/3} \sigma^{-1}, \tag{2.5}$$

where $\rho$ is the physical density, $n$ is the number density and $m$ is the mass of the planetesimal. We conclude that the relative ratio of binary-induced collisions to direct collisions is proportional to $a^2$. Thus, the effect of binary depends strongly on $a$ and can be very large in the TNO region.

From eq.(2.4) we can also estimate how the formation rate of binaries changes as the mass of the planetesimals changes. If we assume that number density $n$ and surface number density $N_{\text{sf}}$ to be $n \sim N_{\text{sf}}/\sigma$ (assuming angular velocity $\Omega = 1$) and $\sigma \propto m^{1/3}$ (assuming $e \propto h = (m/3M_\odot)^{1/3}$), the formation rate $\dot{n}_{b,s}$ is proportional to $m^{-2}$. On the other hand, the dependence of the collision rate is $m^{-5/3}$ (eq.2.5). This simple estimate leads to an conclusion that when planetesimals are small, numerous binaries may be formed. If so, the accretion time scale of the planets should be strongly affected by binaries. If we assume a disk of MMSN, eq.(2.4) tells us that the hard binary formation rate would be $\sim 0.1$ per orbital period at 30 AU in a ring with $\Delta a = 0.3$AU, planetesimal mass of $10^{23}$g and $e \sim 10^{-3} \sim 3.9h$. On the other hand, the rate of direct collision is $\sim 1$ from eq.(2.1). Softer binaries can contribute to the collision as well. Therefore, the number of binary-induced collisions is comparable to that of direct collisions for the case of massive planetesimals of $10^{23}$g, and can be larger for small mass. In the next section, we investigate such effect quantitatively, using direct $N$-body simulation.

# 3 Model

## 3.1 Numerical Setup

We consider particle swarm orbiting around the Sun and investigate their orbital evolution. The equation of motion of particles is

$$\frac{d\boldsymbol{v}_j}{dt} = -\frac{GM_\odot}{|\boldsymbol{r}_j|^3}\boldsymbol{r}_j - \sum_{j \neq k} \frac{Gm_{\mathrm{p},j}}{|\boldsymbol{r}_j - \boldsymbol{r}_k|^3}(\boldsymbol{r}_j - \boldsymbol{r}_k) \quad (3.1)$$

where $\boldsymbol{r}_j$ is the heliocentric position vector, $\boldsymbol{v}_j$ is the velocity vector and $m_\mathrm{p}$ is the mass of particle $j$, and $G$ and $M_\odot$ are the gravitational constant and the mass of the Sun, respectively. The first and second terms are the gravity of the central star and the mutual gravity between the particles, respectively. Since the total mass of the planetesimals (and the protoplanets) is $\sim 10^{-6}$ times the mass of the central star, we neglect the indirect term. We use the 4th order Hermite scheme (Makino and Aarseth 1992) for the orbital integration. The mutual gravity term is calculated using GRAPE-6 (Makino et al. 2003) and GRAPE-6A (Fukushige et al. 2005). We also take into account the particle accretion following the treatment in the previous works (e.g. Kokubo and Ida 1995,1998). We use the perfect accretion model, in which we regard two planetesimals to have merged when the distance between them becomes less than the sum of their radii. Physical radius of a planetesimal is determined by its mass $m_\mathrm{p}$ and the solid density $\rho_\mathrm{p}$ as

$$R_\mathrm{p} = f\left(\frac{3}{4\pi}\frac{m_\mathrm{p}}{\rho_\mathrm{p}}\right)^{1/3} \quad (3.2)$$

We use parameter $f$ to control the collision rate. In simulations with large $f$, collisions are more frequent. We assume $\rho_\mathrm{p} = 3\mathrm{g/cm}^3$ in this paper. The solid density of the planetesimals in TNO is not well understood, and it might be more reasonable to use smaller size. However, in order to compare our result with that of the accretion in terrestrial planet region, we adopted the typical solid density in terrestrial region, which is $3\mathrm{g/cm}^3$. For boundary condition, we used free boundary conditions.

## 3.2 Simulation Parameter

We consider a disk consists of planetesimals around the Sun. The disks have a finite width of $\Delta a_\mathrm{in}$ which is proportional to the semimajor axis from the Sun $a_\odot$. The ratio $\Delta a_\mathrm{in}/a_\odot$ is set to 0.01. The initial mass of planetesimals is equal ($m_p = 10^{23}$g). Parameters are $a_\odot$, planetesimal surface density $\Sigma$, and physical radius enhancement factor $f$. Model parameters of the simulations are listed in Table 1. These values are chosen in order to keep the number of planetesimals manageable (largest $n_{\rm init}$ is 92442). As a result, the planetesimal mass is larger than a realistic value. As we have seen in section 2, the relative importance of three-body binary formation is larger for smaller mass. Therefore, by using relatively large planetesimals, we consider the lower limit of the effect of the three-body binaries.

We consider three different sets of initial models. In the first set, we changed the parameter $f$ (eq.3.2), in order to compare the values of the binary formation frequency for different values of $R_p/r_H$, while keeping all other parameters unchanged. In the second set, we changed the solid surface density at fixed semimajor axis in order to see the surface density dependence of binary formation frequency. Finally, in the third set, we changed the semi-major axis $a_\odot$ following the minimum mass solar nebular model (MMSN) surface density. The surface density that follows MMSN is given by

$$\Sigma_0 = 10 \eta_{\rm ice} \left(\frac{a_\odot}{1\text{AU}}\right)^{-3/2} (\text{g cm}^{-2}), \quad (3.3)$$

where $\eta_{\rm ice}$ is an enhancement factor that expresses ice condensation; $\eta_{\rm ice} = 1.0$ for $a_\odot < 2.7$ and $\eta_{\rm ice} = 4.0$ for $a_\odot > 2.7$. Note that eq(3.3) is slightly larger than that of the original MMSN (Hayashi 1981). Surface density enhancement $\Sigma/\Sigma_0$ is a parameter in simulations of Set 2, which is listed in Table 1 as $g$. In set 1, we performed 6 simulations with different $f$. In set 2, 4 simulations with different $g$ were performed. In set 3, 5 simulations with different semimajor axis of the planetesimal rings were performed. The azimuthal distribution of the planetesimals is chosen randomly. When a particle is bound to another particle initially, we removed that bound particle and added a new one generated randomly. Hence, there is no bound pair initially.

The initial eccentricities and inclinations of planetesimals are given by Rayleigh distribution with dispersion of $\langle e^2 \rangle^{1/2} = 2\langle i^2 \rangle^{1/2} = 0.001$ (Ida and Makino 1992a). Since the gas is still abundant in the planetesimal accretion stage, random velocity is damped by gas to $\langle e^2 \rangle^{1/2} \sim 5h$ (Kokubo and Ida 2002), where $h = ((m_p)/(3M_\odot))^{1/3} \sim 2.5 \times 10^{-4}$. The value of $\langle e^2 \rangle^{1/2}$ we choose does not differ significantly from this value. Though this equilibrium value might depend on the distance from the sun, we ignore that because of the reasons we discussed in section 2.

## 3.3 Definition of a Binary

As shown in later sections, in our simulations, there formed many pairs of particles which seem to be bound by their mutual gravity. Since the particles are not in free space, but orbiting around the Sun, the binding energy of particles is not suitable for the definition of such "binary". In a planetesimal disk, with a Hill approximation, Jacobi energy of two planetesimals is defined as

$$U = \frac{1}{2}\left(\dot{x}^2 + \dot{y}^2 + \dot{z}^2\right) - \frac{3}{2}\Omega_k^2 x^2 + \frac{1}{2}\Omega_k^2 z^2 - \frac{2Gm_p}{r} + \frac{9}{2}r_H^2\Omega_k^2, \qquad (3.4)$$

where $x$ are the relative position and velocity in barycentric, rotating coordinate, $r_H$ is the mutual hill radius $(2m_p/3M_\odot)^{1/3}a_c$ ($a_c$ is semimajor axis of center of mass), $\Omega_k$ is the angular velocity of the center of mass, and $r$ is the distance between the two planetesimals. The last term is added so that $U$ at Lagrange points $L_1$ and $L_2$ is 0. When $U < 0$ and $r < r_H$, two particles are clearly bound. However, there are many unclear cases. Not only planetesimal pairs with $U < 0$, but also very soft temporareley captured pairs, with $U \sim 0$ (Iwasaki and Ohtsuki 2007) can contribute to collision enhancement. We do not want to neglect such pairs. Hence, $U$ can not be used for a criterion of a pair we are looking for. In addition, quasi-satellites are likely to be formed in planetesimal disk. Quasi-satellites are ones that are not bound but orbits around each other with separation of $\sim r_H$. Such pairs can also form binaries by three-body encouters. We found that that quasi-satellites tend to become bound eventually and increase the collision rate. In this paper, we use the criterion $a_b < 2r_H$, in order not to miss the soft temporarely captured pairs and to include quasi-satellites. We define the pair satisfying this criterion as a "binary "and use the term in this paper.

# 4 Results

## 4.1 Examples of binary formation and evolution

First we show trajectories of particles for typical binary formation events. Figures 2 and 3 are examples of how a binary is formed by multi-body encounter and how it ends by collision. These are taken from model R30AUf1.0. Each plot corresponds to 4(e) and 4(g), respectively. One of the planetesimals of the binary is plotted as a black dot, which is the center of the coordinate. Another participant which eventually become the other component of the binary is shown in red dots. Other planetesimals which perturb the orbits of the first two particles are shown in blue, green, purple and yellow. The dots are plotted every $1/(2\pi)$years. The orbits are plotted in the rotating frame. In fig.2, the red particle which forms a binary until the collision enters the plot at 122.1. years. Encounter with the blue particle, green particle and purple particle takes away angular momentum between red and black planetesimals. The orbit of the red planetesimal results in highly eccentric orbit around the black planetesimal. The perturbation from the yellow particle, which enters the plot at 302 years, makes the red and black planetesimals collide. In the second sample (fig.3), the red particle encountered the blue particle close to the black particle at about 100 years. It makes the red particle weakly bound to the black particle. The orbit of the binary reaches about 0.005 AU. The perturbation from green, purple, light blue and yellow particle makes the binary's semimajor axis smaller. Another planetesimal was about to enter the plot at 626.7 years and made the red and black planetesimal collide. Binary formation events like these are not rare as shown in Table 1. The total number of binaries is about several hundred in 1000 years. The number of binaries with $a_\mathrm{b} < r_\mathrm{H}$ is $\sim$ 10-20 (in standard R30AUf1.0 case). These binaries contribute to the collision rate.

Figure 4 shows the time evolution of the separations between two bodies that eventually collide in model R30AUf1.0. The total number of collisions in this run is 20 (see Table 1). This figure shows all collisions in which the colliding particles satisfy the criterion $a_\mathrm{b} < 2r_\mathrm{H}$ just before the collision. The semi-major axis of each pair is also plotted in red dots. Time evolution of Jacobi energy of each pair is drawn in the lower box. Binaries or at least temporarily captured pairs have significant effect on how collisions occur and how many of them occur. In almost half of the collisions, the collision participants were bound in some way. Cases (e),(g) and (h) show clear sign of binary orbits. Cases (f) and (i) show weak sign of bound orbits. We will see how the number of collisions changes when the initial model parameters such as the radial enhancement factor, the number density and the distance from the Sun are changed in the next subsection. The collisions which took place when $t$ is small are hard to define as a "binary", and similar to direct collision. Formation of such quasi-binary may depend on initial condition. We need more sample to argue statistically e.g. simulations with same initial condition with different angular distribution. Such investigation is left for future study.

## 4.2 Collision Probability

Figure 5 shows the total number of collisions for simulations in Set 1, where the physical radius of planetesimals were changed and all other parameters were kept the same. The solid line shows the total number of collisions $n_c$, and the dashed curve the number of collisions in which binaries are not involved ($n_c - n_{b,c}$). The number $n_{b,c}$ is defined as the number of collisions in which one of the two colliding planetesimals was a member of a binary in the snapshot output just before the collision. The dotted line in Fig. 5 shows a theoretical estimate for the number of collisions, obtained for simple model of direct collision of two particles. If we neglect binaries, collision probability $P$ (number of collision per unit time for one planetesimal) can be calculated using two-body approximation (Ida and Nakazawa 1989, Greenzweig and Lissauer 1990) as:

$$P \simeq \frac{n_s}{2 a_\odot i} \pi R_P^2 \left(1 + \frac{v_{esc}^2}{v_{rel}^2}\right) v_{rel}, \qquad (4.1)$$

where $n_s$ is the surface number density of the planetesimal, $v_{esc}$ is the escape velocity and $v_{rel}$ is the relative velocity between the planetesimals. Since the eccentricity of the planetesimals are small ($\sim 10^{-3}$) in our simulations, gravitational focusing is effective ($v_{esc} > v_{rel}$) and $v_{rel}$ can be written as the velocity dispersion of the field bodies ($v_m \sim \sqrt{e_m^2 + i_m^2} v_{kep}$, $e_m$ and $i_m$ are the RMS eccentricity and inclination), and the first term in the parenthesis of Eq.(4.1) can be neglected. Using collision probability with two-body approximation described above, we can estimate the number of collisions which are not binary originated for each value of $r_H/R_P$. We can easily see that $P \propto R_\mathrm{p}/r_\mathrm{H}$. Velocity dispersion of the planetesimals in each simulation in Set 1 is shown in Fig.6. Time evolution of RMS eccentricity and inclination are plotted. Theoretical estimate of number of collision is derived by integrating $P$ for 1000 years using RMS eccentricity and inclination in each simulation. As shown in figure 6, all the simulations in Set 1 exhibits almost the same time evolution for eccentricity and inclination. Therefore, the difference of the number of collisions should come solely from the difference of the size of the planetesimals. We can see that $n_\mathrm{b,c}$ is significant compared to total collision number $n_\mathrm{c}$. This means that binary formation affects the collision mechanism as least in the outer region of the disk. In addition, we can see that the slope of the number of collisions in which binaries are not involved shows a reasonable agreement with the theoretical estimate.

Figure 7 shows the ratio of collisions in which the binaries took part in to the total number of collisions, as a function of $r_\mathrm{H}/R_\mathrm{p}$. The figure shows the number of binary related collisions are $\sim 1/2 - 1/3$ times the total number of collisions in each run. As shown in table 1, the numbers of binaries formed are about the same. However, the number of collisions increases as $r_\mathrm{H}/R_\mathrm{p}$ decreases, and $n_\mathrm{b,c}$ increases as well. The parameter $r_\mathrm{H}/R_\mathrm{p}$ corresponds to the displacement from the Sun. Small $r_\mathrm{H}/R_\mathrm{p}$ corresponds to inner region of the disk and large $r_\mathrm{H}/R_\mathrm{p}$ corresponds to the outer region of the disk. Collision probability increases with $f$. The binary formation rate does not depend on $r_\mathrm{H}/R_\mathrm{p}$. The reason why $n_\mathrm{b,c}$ is larger for large $f$ is simply that the collision time scale is smaller for large $f$. The reason why no equal mass binary is found in inner region is that collision frequency is high in the inner region and no equal mass binary is left. In the outer region, the binaries can survive because of the low collision frequency.

## 4.3 Effect of the number density

Figure 8 shows the number of collisions $n_\mathrm{c}$ as the function of the surface number density $\Sigma$ normalized by the MMSN value $\Sigma_0$. The dotted line shows the analytical estimate based on eq.(4.1). As in the previous section, we plot the total number of collisions occurred in 1000 years (solid line). We can see that the number of collisions observed is a few times larger than the theoretical estimate. The actual number of collisions in which binaries are involved is $n_\mathrm{b,c}$ (which is not drawn in the figure). We can see that the number of direct collisions (binaries are not involved in, $n_c - n_{b,c}$ which is plotted as a dashed line) is closer to the simple theoretical estimate. Note that $n_c$ and $n_c - n_{b,c}$ are not plotted if they are equal to zero.

## 4.4 Distribution of binaries

In this subsection we investigate how the distribution of binaries (total numbers and Jacobi energies) changes in time. Time evolution of binary distribution of simulations of Set 1 is shown in Figs. 9 (a)-(f). As the absolute value of $U$ increases, the semimajor axis of the binary decreases and the binary becomes harder. Number of binaries with smaller $U$ increases as time goes on. The increase is larger in simulations with smaller $f$. In fig. 9(c), which shows the standard case, dashed line is plotted to describe the cumulative number of binaries with $a_b < r_H$ at 1000 years. Tendency that the binaries with smaller $U$ increases is the same as the result obtained by setting the criterion to be $a_b < 2r_H$.

Figure 10 plots the total increase of the number of binaries with $U < 0$ as a function of $r_H/R_p$. There is a trend that the increase is large for smaller $R_p$. The change in the physical radius of the planetesimals should not affect the binary formation. However, the collision probability increases as $r_H/R_p$ decreases, resulting in larger number of binaries in cases with large $r_H/R_p$.

It is highly possible that the random velocity of planetesimals are small when they are formed. When the RMS eccentricity and inclination are larger than that of R30AUf1.0(standard case), smaller number of binaries are formed (fig.11). Hence, smaller RMS eccentricity and inclination should enhance the binary formation and collision rate.

Binary formation is affected by surface density, as well as $r_H/R_p$. Figures 12 shows the binary distribution in terms of Jacobi energy for simulations in Set 2. The solid lines show the distribution after 100 years. In (a) and (b), the dotted lines show the distribution after 1000 years. Since the encounter rate of planetesimals is smaller in simulations with smaller $\Sigma$, we plot the distribution at $t = 3000$ years in (c)($\Sigma = 0.3\Sigma_0$) and at $t = 10000$ years in (d)($\Sigma = 0.1\Sigma_0$) with dotted lines. In R30AUg3.0, the number of tight binaries increases after 1000 years. As the surface density decreases, number of binaries with small jacobi energy decreases, and the increase of such binaries also becomes smaller. In R30AUg0.1, only few binary with negative

Jacobi energy exists even after 10000 years.

Now we investigate the "hard" binary formation rate. Here we define a "hard" binary as a binary with $a_b < r_H$ to compare our results with eq.(2.4). Three colored curves in Figure 13 show the binary formation rate per year for binaries with $a_b < r_H$ as a function of $\Sigma/\Sigma_0 (= g)$. We define the formation rate as

$$\left\langle \frac{n_b(t + \Delta t) - n_b(t)}{\Delta t} \right\rangle. \tag{4.2}$$

Since some binaries have very short lifetime, the above formation rate depends on the value of $\Delta t$. That is why we used several different values of $\Delta t$. The formation rate dependence is close to the square of $\Sigma$, independent of the choice of $\Delta t$. Since the planetesimal mass is the same in all simulations, the surface density corresponds to the number density of planetesimals. The dependence on $\Sigma$ of binary formation rate (per year per volume) for star cluster (Eq.(2.4)) and that of SS08(per year per volume) is $\Sigma^3$. Our simulation result does not agree with the result of binary formation rate in star cluster nor SS08. The reason is that in star clusters, the binaries with large binding energy ($a_b < 0.1 r_H$) are counted, while we counted relatively loose ($a_b < 2 r_H$) ones. Binaries can be stable with such binding energy in planetesimal disks. The binary formation rate when $g = 1$ is consistent with SS08. However, the dependence does not agree well with the prediction. In SS08, velocity dispersion is smaller than that of our simulations. This may cause the inconsistency with our results and that of SS08. A detailed comparison is left to future study.

## 4.5 Dependence on the Distance from the Sun

Now we show the result of simulations of Set 3. Surface density of the planetesimal ring of each run in this set is consistent with MMSN. In this case, the semimajor axis and the surface density vary simultaneously. First we focus on the number of collision, as we did in sections 4.2 and 4.3. Figure 14 shows the number of collisions plotted against the semimajor axis in 1000 years. Not only in the outer region but also in the inner region of the disk, the number of collision is significantly larger than the analytical estimate which does not take into account the effect of binary formation. Since the initial planetesimal rings have different semimajor axes ($a_\odot$) and different surface density (following MMSN), orbital period of each simulation and the encountering rate varies. The fraction of $n_{b,c}$ ($n_{b,c}/n_{col}$) is larger in the outer region (Fig. 15). Almost half of the collisions are binary related. This result implies that binary formation enhances the planetesimal collision by a factor of a few.

Figure 16 shows the cumulative number of binaries as a function of Jacobi energy, as in Figs. 9 and 12. In disks with small semimajor axis, binaries with $a_b < 2r_H$ can be formed but destroyed quickly afterwards. Hard binaries tend to become harder in the region with large semimajor axis. Our result shows that binaries are also formed in inner region of the disk, while no equal-mass binary has been observed in inner region of the disk. This is because of the short orbital period (hence fast orbital evolution) in the inner region than in the outer region. Also, collision takes place more easily because the $r_H/R_p$ is small. Planetesimal encounters and collisions destroy the binaries in the inner region and they cannot survive. Hence, they are not observed so far.

# 5 Discussion

## 5.1 Effect on the Accretion Time Scale

The accretion time scale in terrestrial planet region is investigated by both analytical and numerical methods (e.g. Kokubo and Ida 2002). Their estimated time scale is based on two body approximation. The collision probability is given by Eq. (4.1). As in the previous section, binary formation affects the collision probability. If the collision probability differs significantly from that derived assuming two body approximation, the accretion time scale of planetesimals is affected as well.

In the region where binary formation is frequent, binary-induced collisions should be taken into account to the estimate of the collision frequency. Binary originated collision rate can be approximated by the binary-single planetesimal encounter rate. Using the result of R30AUf0.3, number of binary formed is $\sim 200$. Binary-single planetesimal encountering rate that leads to merger is estimated by Hut and Inagaki (1985) as

$$n<\sigma V> N_{B2} = \frac{10^{-8} F^{7.0.6} N_{B2}}{t_r \ln(0.5N)} \left(\frac{m_p}{10^{23} \text{g}}\right)^{-1} \left(\frac{R_p}{200 \text{km}}\right) \left(\frac{e}{10^{-3}}\right)^2 \quad (5.1)$$

where $N_{B2}$ is the number of binaries, $t_r$ is relaxation time and $N$ is the number of planetesimals. Parameter $F$ is ratio $a_b/(4R_p)$. Assuming $a_b \sim r_H$, $F \sim 100$. Hence,

$$n<\sigma V> N_{B2} = \frac{10^{-6} N_{B2}}{t_r \ln(0.5N)} \left(\frac{m_p}{10^{23} \text{g}}\right)^{-1} \left(\frac{R_p}{200 \text{km}}\right) \left(\frac{e}{10^{-3}}\right)^2. \quad (5.2)$$

If we assume the number of planetesimal binaries $N_{B2} \sim 10^3$ and $t_r$ to be Kepler time, and the number of planetesimals is $N \sim 30000$ the encountering rate would be $\sim 10^{-3}$ per binary $t_r$. Since the number of binary is about $10^3$ after $10t_r$, the number of encounters would be estimated as order of $\sim 1$ which is consistent with our simulation result. Together with the collision probability Eq. (4.1), new collision probability would be enhanced by a factor of a few of that of two body approximation. At the early stage of the planetary formation, the abundance of the planetesimals might have been larger than that of now. The number density (or surface density) of the planetesimals in the outside region could be larger by several factor. In addition, random random velocity of the planetesimals could be smaller (Weidenschilling 2002). Such condition enhances the planetesimal binary formation, and results in increase of accretion rate of planetesimal by more than a few. In situ formation of dwarf planets may be able to be explained considering the binary formation. However, longer simulation is necessary to figure out the precise factor. Such simulations are left to future study.

## 5.2 Summary

We performed N-body simulations of planetesimal rings with various semimajor axis, surface density and physical radius. The main results are the following.

- The number of collisions in which binary planetesimals are involved is pretty large, at least half of the all collisions. First, gravitationally bound pairs of planetesimals are formed. Then surrounding planetesimals encounter with the bound pairs, making them collide. There is an addition of binary originated collision. When $r_\mathrm{H}/R_\mathrm{p}$ is small, collision takes place before binary forms.

- Number of binaries with negative Jacobi energy increases with time. These "hard" binaries encounters other planetesimals and eventually collide. If we assume that the encounter rate is the collision rate, the planetesimal accretion rate becomes about $\sim 2-3$ times the analytical estimate that does not include the effect of binary formation.

- As the number density of the planetesimals (or the surface density of planetesimal disk) increases, the binary formation rate increases. The order of binary formation rate when $g=1$ was consistent with SS08. However, the velocity dispersion and hardness of the binaries we considered are different from those of SS08 and star clusters. The comparison is left to future study.

In our Solar system, planetesimal binaries with large $a_\mathrm{b}$ and large $e_\mathrm{b}$ are observed in TNO region. These binaries are not found in inner region of the system. Our simulation results show that hard binaries are more easily formed in regions with large $r_\mathrm{H}/R_\mathrm{p}$, which is larger in outer region of the disk. Hence, our simulation results are consistent with the observation. In addition, our results show that binary formation tends to shorten the accretion time scale in TNO region.

All our simulations started with equal mass planetesimals. In order to investigate if binary formation by collision and exchange reaction (Funato et al. 2004) takes place or not, fragmentation has to be modeled as well. Longer simulations has to be carried out to investigate the runaway accretion and oligarchic growth phase. These effects will be investigated using more realistic model in the forth coming papers.

## Acknowledgment

This work is supported by JSPS Research Fellowship (19.3622). Some of the numerical simulations were carried out on GRAPE system at Center for Computational Astrophysics (CfCA) of National Astronomical Observatory of Japan.

## Table 1

## Simulation List

### (Set 1)

| Simulation | $n_{\text{init}}$ | $\Delta a_\odot$ (AU) | $f$ | $g$ | $n_c$ | $n_{b,c}$ | number of binaries($t = 1000$years) |
|---|---|---|---|---|---|---|---|
| R30AUf0.1 | 30812 | 0.3 | 0.1 | 1.0 | 3 | 2 | 365 |
| R30AUf0.3 | 30812 | 0.3 | 0.3 | 1.0 | 6 | 3 | 372 |
| R30AUf1.0 | 30812 | 0.3 | 1.0 | 1.0 | 20 | 9 | 316 |
| R30AUf3.0 | 30812 | 0.3 | 3.0 | 1.0 | 50 | 20 | 381 |
| R30AUf10.0 | 30812 | 0.3 | 10.0 | 1.0 | 152 | 64 | 382 |
| R30AUf30.0 | 30812 | 0.3 | 30.0 | 1.0 | 434 | 154 | 340 |

### (Set 2)

| Simulation | $n_{\text{init}}$ | $\Delta a_\odot$ (AU) | $f$ | $g$ | $n_c$ | $n_{b,c}$ | number of binaries |
|---|---|---|---|---|---|---|---|
| R30AUg0.1 | 3080 | 0.3 | 1.0 | 0.1 | 0 | 0 | 9(10000yrs) |
| R30AUg0.3 | 9243 | 0.3 | 1.0 | 0.3 | 0 | 0 | 48(3000yrs) |
| R30AUg1.0 | 30812 | 0.3 | 1.0 | 1.0 | 20 | 9 | 316(1000yrs) |
| R30AUg3.0 | 92442 | 0.3 | 1.0 | 3.0 | 150 | 80 | 1485(1000yrs) |

### (Set 3)

| Simulation | $n_{\text{init}}$ | $\Delta a_\odot$ (AU) | $f$ | $g$ | $n_c$ | $n_{b,c}$ | number of binaries($t = 1000$years) |
|---|---|---|---|---|---|---|---|
| R3AU | 9743 | 0.03 | 1 | 1 | 164 | 34 | |
| R5AU | 12570 | 0.05 | 1 | 1 | 114 | 39 | |
| R10AU | 17789 | 0.1 | 1 | 1 | 49 | 22 | 51 |
| R20AU | 25159 | 0.2 | 1 | 1 | 22 | 9 | 215 |
| R30AU | 30812 | 0.3 | 1 | 1 | 20 | 9 | 316 |

Parameters and number of collisions of the simulations are listed. $n_{\text{init}}$ is the initial number of the planetesimals. $f$ is the physical radius enhancement as in Eq.(3.2). $g$ is the surface density enhancement $\Sigma/\Sigma_0$. $n_c$ is the total number of collision took place in 1000 years. $n_{b,c}$ is the number of binary originated collision in 1000 years.

# Figure Captions

**Figure 1**

The ratio of $\sigma_{cross,2}/\sigma_{cross,1}$ is plotted as a function of $a_\odot$(AU). Solid line is plotted assuming a planetesimal binary semimajor axis $a_b = r_H$. Dotted line is plotted assuming $a_b = 0.1 r_H$. We also assumed that the gravitational focusing is dominant $((v_{esc}/v_{rel})^2 > 1)$ and neglect the first term of $\sigma_{cross,1}$.

**Figure 2**

An example of orbital evolution of typical binary formation
(one of the binaries of run R30AUf1.0) which resulted in collision. This collision corresponds to fig.4(e). The origin is located at one of the planetesimals that forms a binary, plotted as a black dot. Another participant of the binary's orbit is plotted in red. The closest approach of purple planetesimal to the black planetesimal is at 159 years. The encounter also affects the orbit of red planetesimal making it bound to the black planetesimal. After the perturbation from the purple planetesimal, the green planetesimal passes by at $\sim$ 165 years. The blue planetesimal purturbs the binary and takes away internal energy between the red and black planetesimals. The closest approach was at 187.6 years. The yellow planetesimal's orbit enters the plot at 302.7 years. It purturbs the red planetesimal and makes the red and black planetesimals collide at 308.4 years.

**Figure 3**

Another example of orbital evolution of typical binary formation
(one of the binaries of run R30AUf1.0) that resulted in collision. The coordinates are the same as fig.2. This collision corresponds to fig.4(g). First, the red planetesimal eventually comes close to the black planetesimal. The planetesimal plotted in blue enters the plot and encounters the red and black planetesimals, (closest at 108.4 years) making the two loosely bound. Next closest encounter (orbit plotted in green) takes place at 298.3 years. The encounter takes away the internal energy of the red and black loosely bound pair. The planetesimal plotted in yellow enters the plot at 361.4 years and leaves the plot at 389.5 years. The planetesimal plotted in yellow also purturbs the red planetesimal's orbit. The red planetesimal's orbit is highly eccentric. When the closest approach takes place between the red and black planetesimals, another planetesimal which is not plotted in the figure, approaches from (x,y)=(-0.01,-0.005). It caused the red and black planetesimals collide at 626.8 years.

**Figure 4**

Time evolution of distance ($r$) between two planetesimals which first form a binary and eventually collide in R30AUf1.0. Total number of collision is 22. Nine of them became binaries and collided. Unit of distance is $r_H$. Binaries' semimajor axes ($a_b(r_H)$) are plotted in red. Jacobi energy evolution of each collision is also plotted below the distance evolution.

**Figure 5**

Number of collisions of simulations in Set 1. X-axis shows $r_H/R_p$ of each simulation. Solid line shows the total number of collision occurred in 1000 years. Dashed line is the number of non-binary related collision ($n_c - n_{b,c}$). The dotted line is the analytical estimate which is calculated using two-body approximation(eq.4.1). Error bar is drawn for each data.

**Figure 6**

Time evolution of RMS eccentricities (upper lines) and RMS inclinations (lower lines) in simulations of Set 1. Every simulation has almost the same $<e^2>^{1/2}$ and $<i^2>^{1/2}$.

**Figure 7**

Number of colliding binaries ($n_{b,c}$) normalized by total number of collision ($n_c$) in simulations in Set 1. Horizontal axis is $r_H/R_H$ of each simulation.

**Figure 8**

Number of collision in terms of surface density $\Sigma/\Sigma_0(=g)$. Dots connected with solid line are total number of collision ($n_c$) in simulations of Set 2. Dots connected with dashed line are the number of non-binary related collision ($n_c - n_{b,c}$). Dotted line is drawn using eq.(4.1). Note that $n_{b,c}$ and $n_c - n_{b,c}$ are not plotted if they are equal to zero.

**Figure 9**

Binary distribution in terms of jacobi energy $U$. The cumulative number is plotted. Dotted line corresponds to distribution at $t = 100$ year and the solid line corresponds to distribution at $t = 1000$ years. (a)-(f) are the results of R30AUf0.1, 0.3, 1.0, 3.0, 10.0 and 30.0, respectively.

Dashed line in (c) shows the cumulative number of binaries with $a_{\rm b} < r_{\rm H}$ in R30AUf1.0.

**Figure 10**

Total increase of the number of binaries after 1000 years as a function of $r_{\rm H}/R_{\rm p}$ in simulations of Set 1.

**Figure 11**

Cumulative number of binary distribution in terms of Jacobi energy $U$. Solid line represents the distribution of R30AUf1.0 at 1000 years. Dashed line and dotted line shows the simulation results starting with twice and four times the random velocity of R30AUf1.0, respectively.

**Figure 12**

Binary distribution in terms of jacobi energy $U$ of simulations in Set 2. The cumulative number is plotted. Dotted line corresponds to distribution at $t = 100$ year and the solid line corresponds to distribution at $t = 1000$ years in (a) and (b). Distribution at 3000 years is shown in (c) and 10000 years in (d). (a)-(d) are the results of R30AUg0.1, 0.3, 1.0 and 3.0, respectively.

**Figure 13**

Binary formation rate(year$^{-1}$) of binaries with $a_{\rm b} < r_{\rm H}$ in simulations of Set 2 . Horizontal axis is $\Sigma/\Sigma_0 (= g)$. The dotted line is analytical line $\propto g^2$ The green line is plotted using snapshots every 17.5 years. The blue line is plotted using snapshots every 50.0 years. The red line is plotted using snapshots every 165 years.

**Figure 14**

Number of collision in terms of semimajor axis $a_\odot$(AU) of the simulations of Set 3. Solid line is total number of collisions, $n_{\rm c}$. The dotted line is plotted using eq.(4.1). The dashed line shows $n_{\rm c} - n_{\rm b,c}$.

**Figure 15**

Fraction $n_{\rm b,c}/n_{\rm c}$ is plotted as a function of $a_\odot$(AU) of simulation of Set 3.

**Figure 16**

Binary distribution in terms of jacobi energy $U$. The cumulative number is plotted. Dotted line corresponds to distribution at $t = 100$ year and the solid lines correspond to distribution at $t = 1000$ years. (a)-(e) are the results of R3AU, R5AU, R10AU, R20AU and R30AU, respectively.

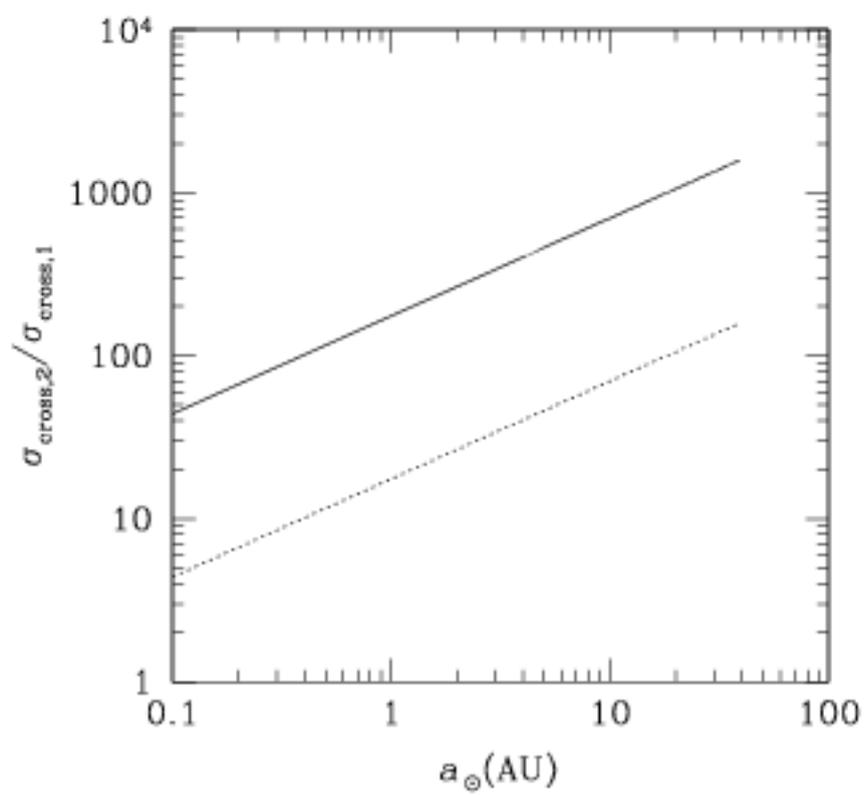

Figure 1: Daisaka.J.K et. al. (2010)

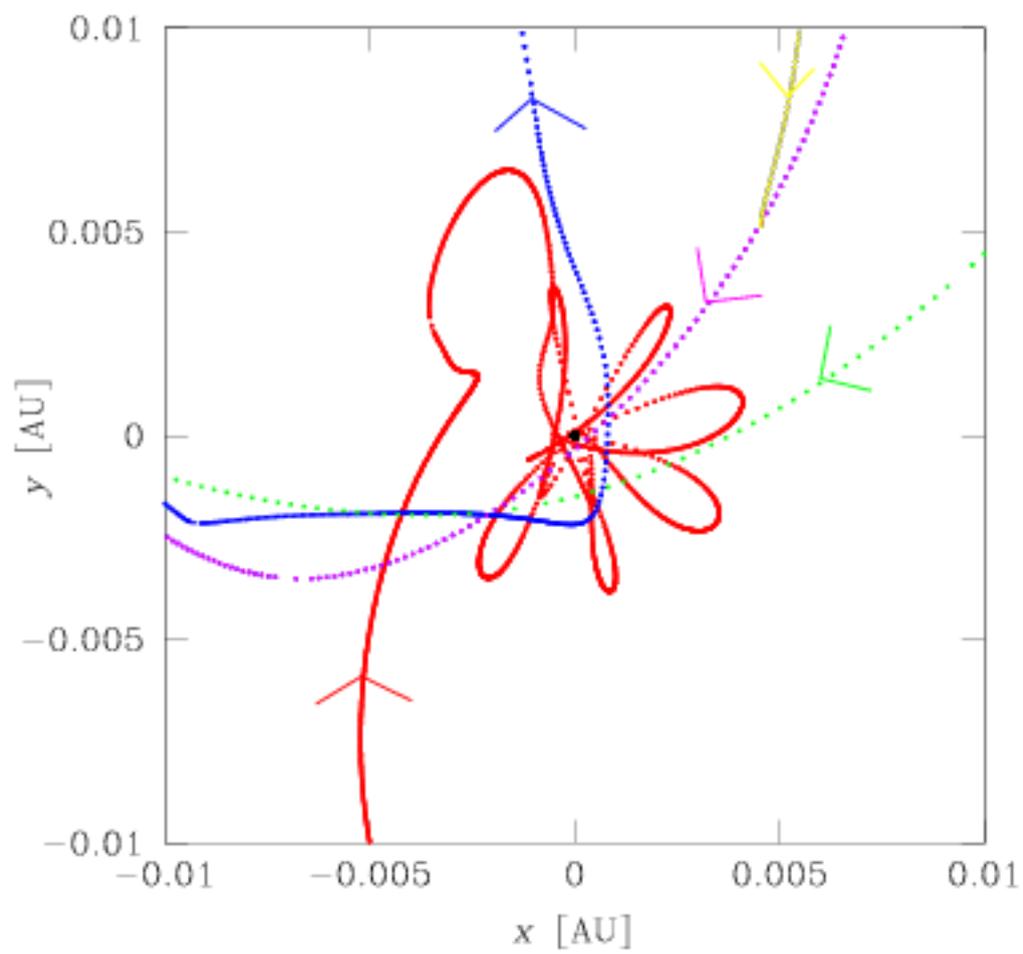

Figure 2: Daisaka.J.K et. al. (2010)

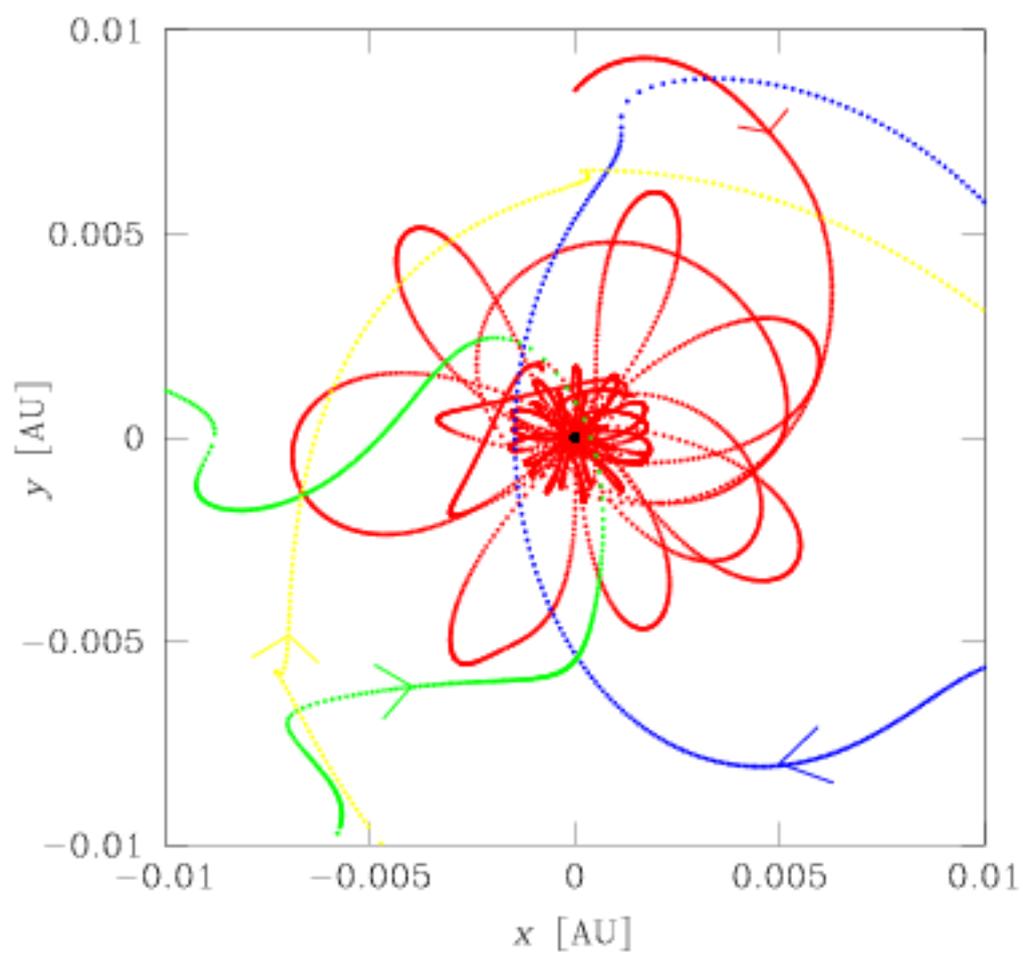

Figure 3: Daisaka.J.K et. al. (2010)

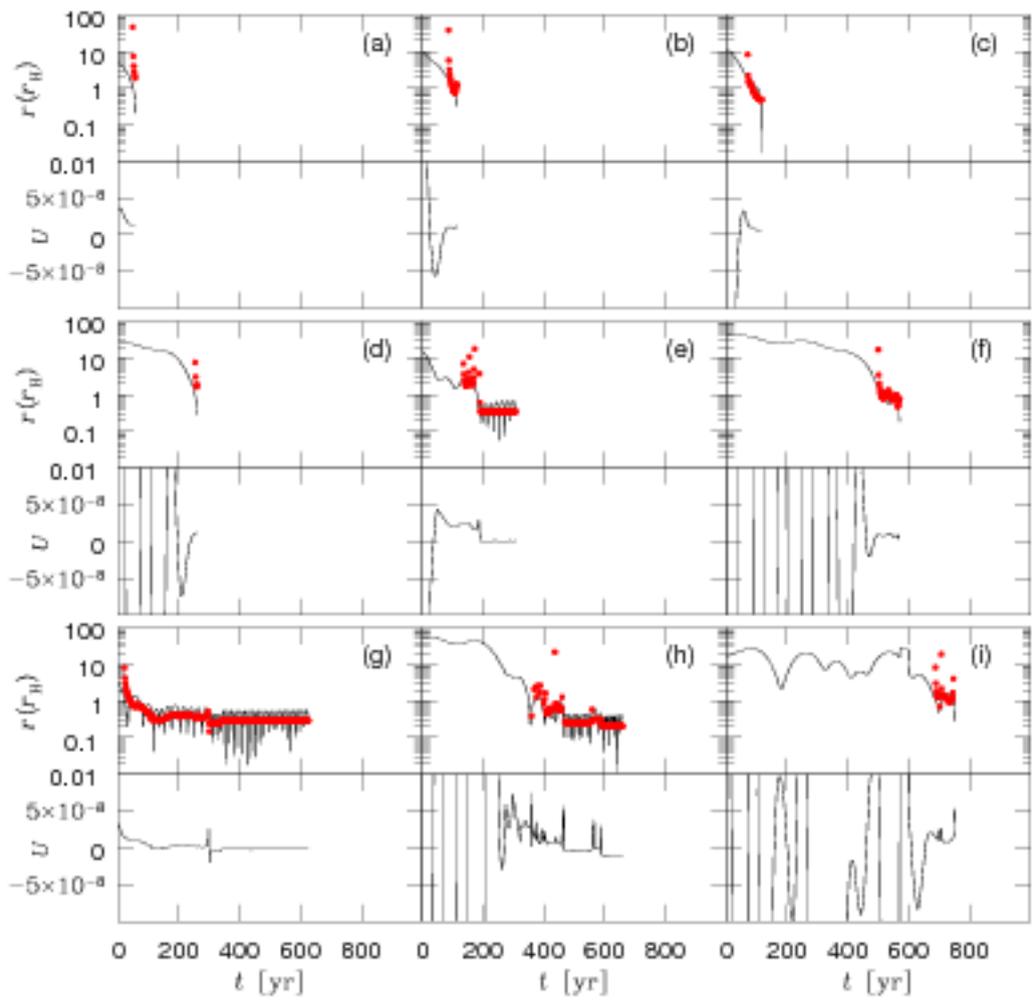

Figure 4: Daisaka.J.K et. al. (2010)

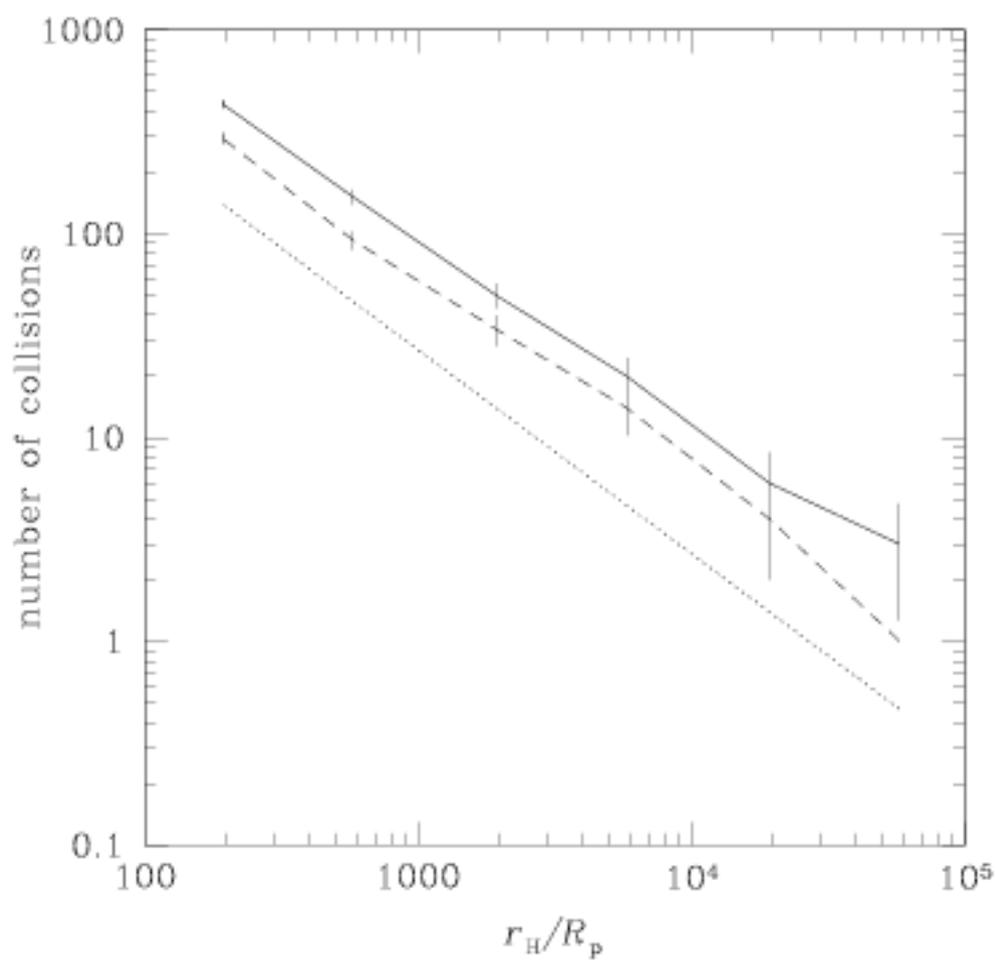

Figure 5: Daisaka.J.K et. al. (2010)

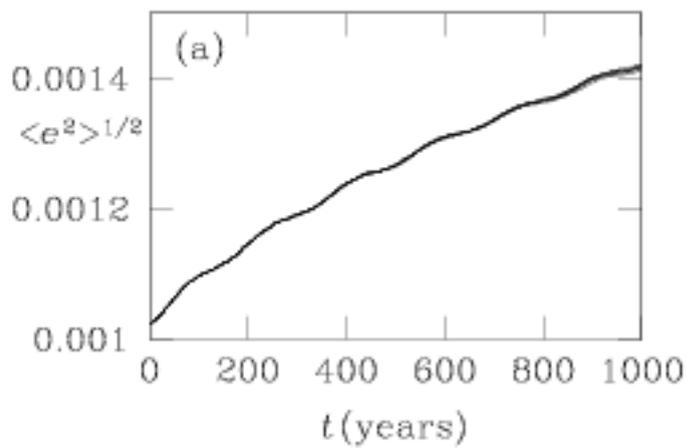

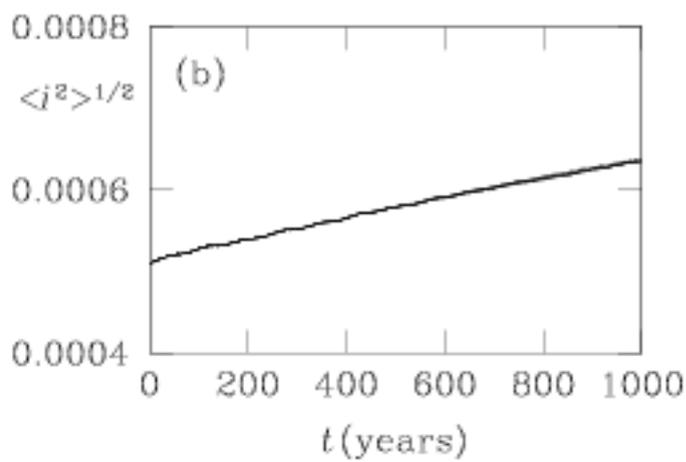

Figure 6: Daisaka.J.K et. al. (2010)

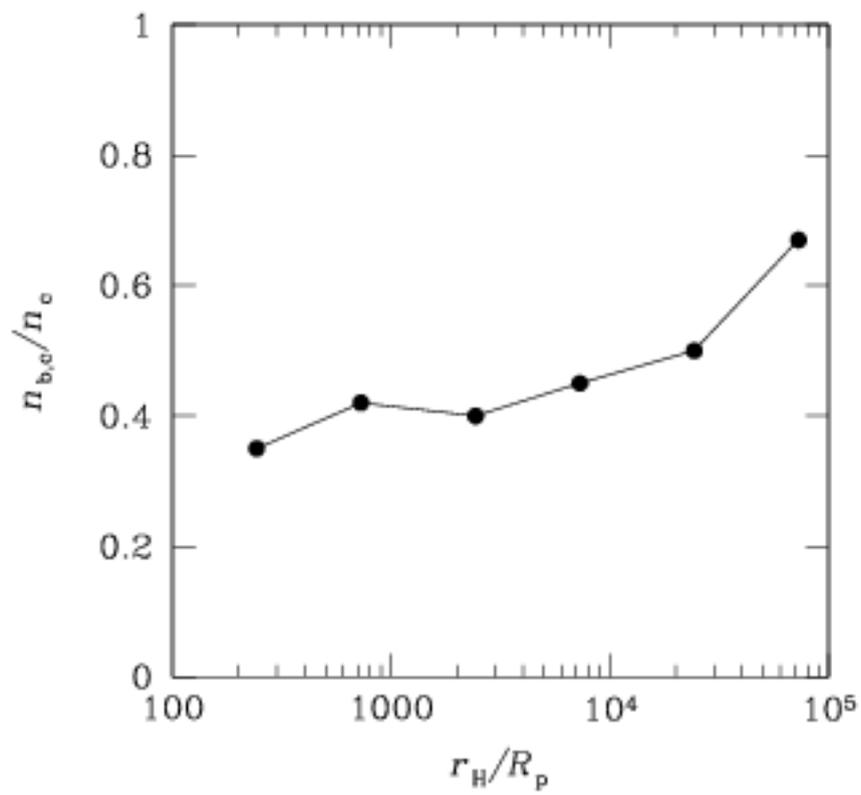

Figure 7: Daisaka.J.K et. al. (2010)

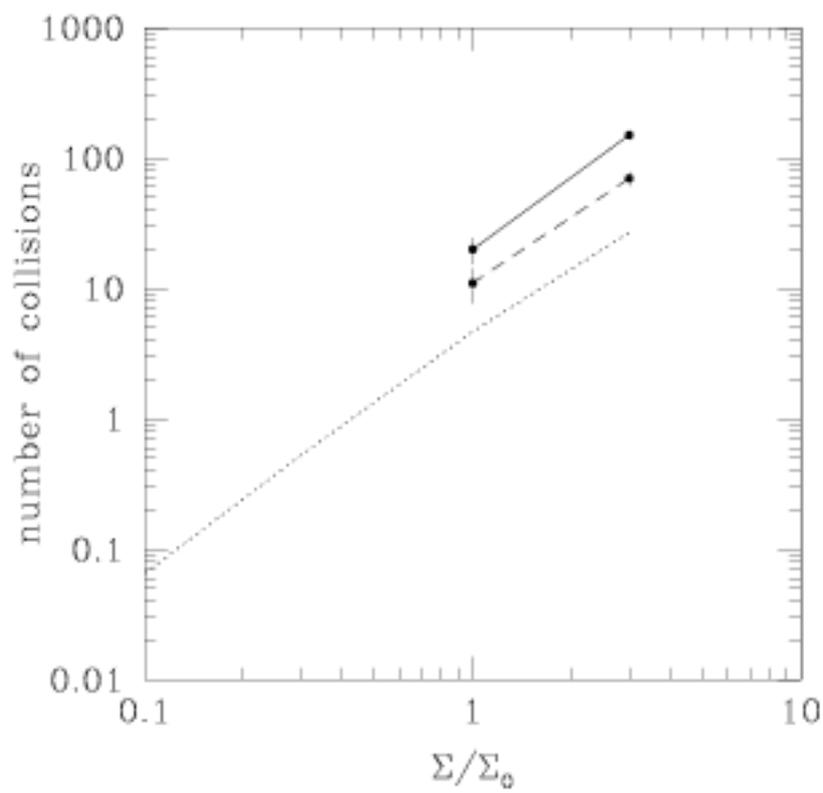

Figure 8: Daisaka.J.K et. al. (2010)

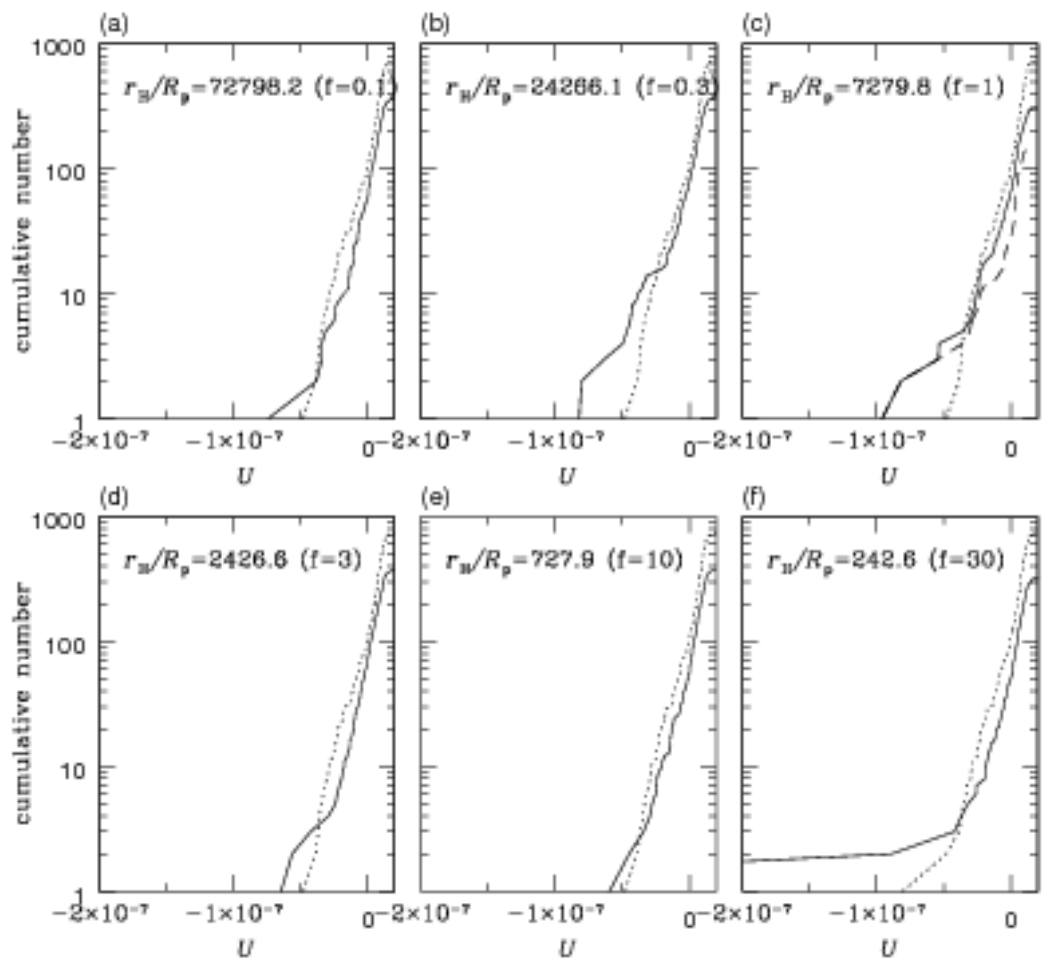

Figure 9: Daisaka.J.K et. al. (2010)

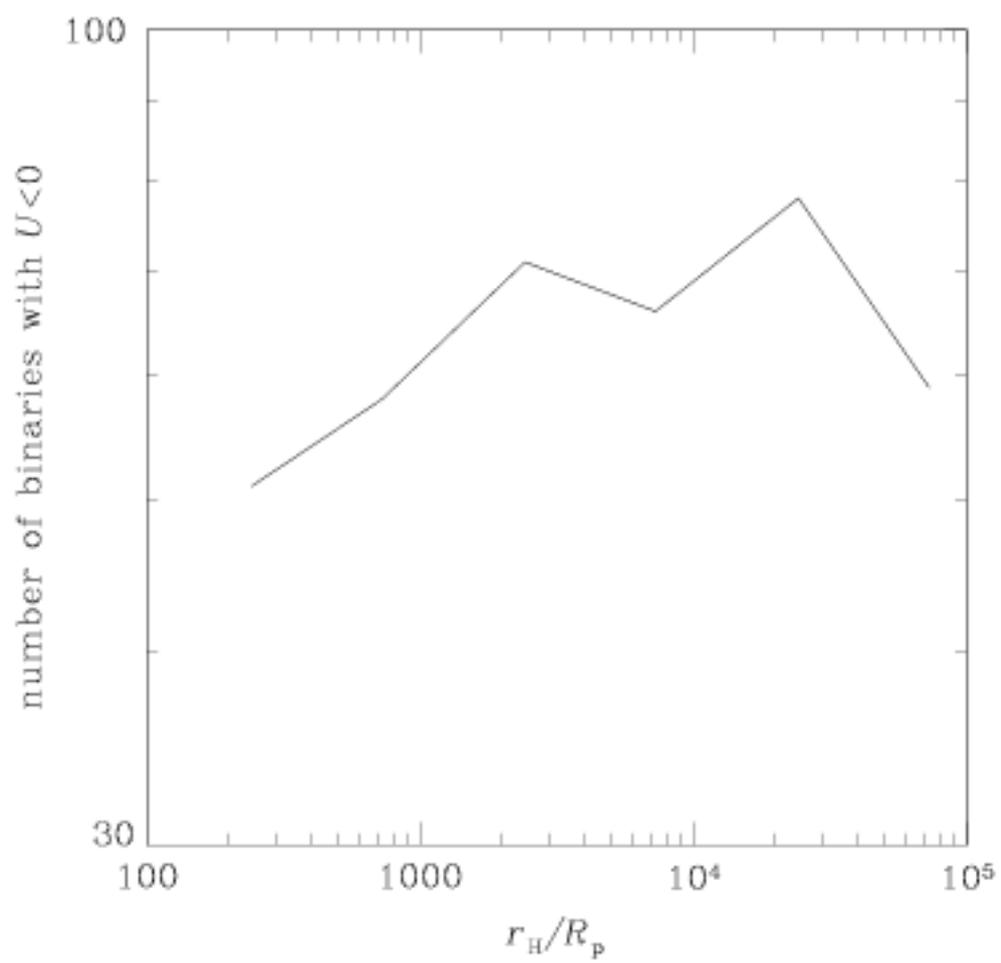

Figure 10: Daisaka.J.K et. al. (2010)

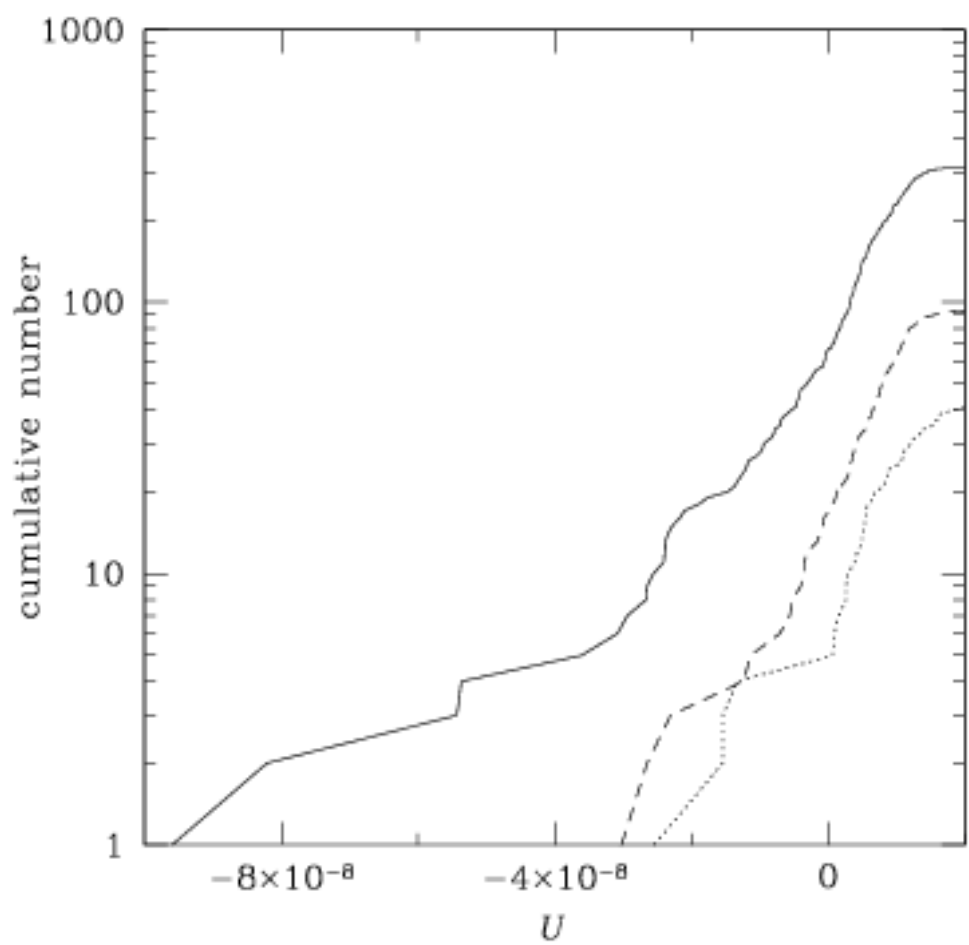

Figure 11: Daisaka.J.K et. al. (2010)

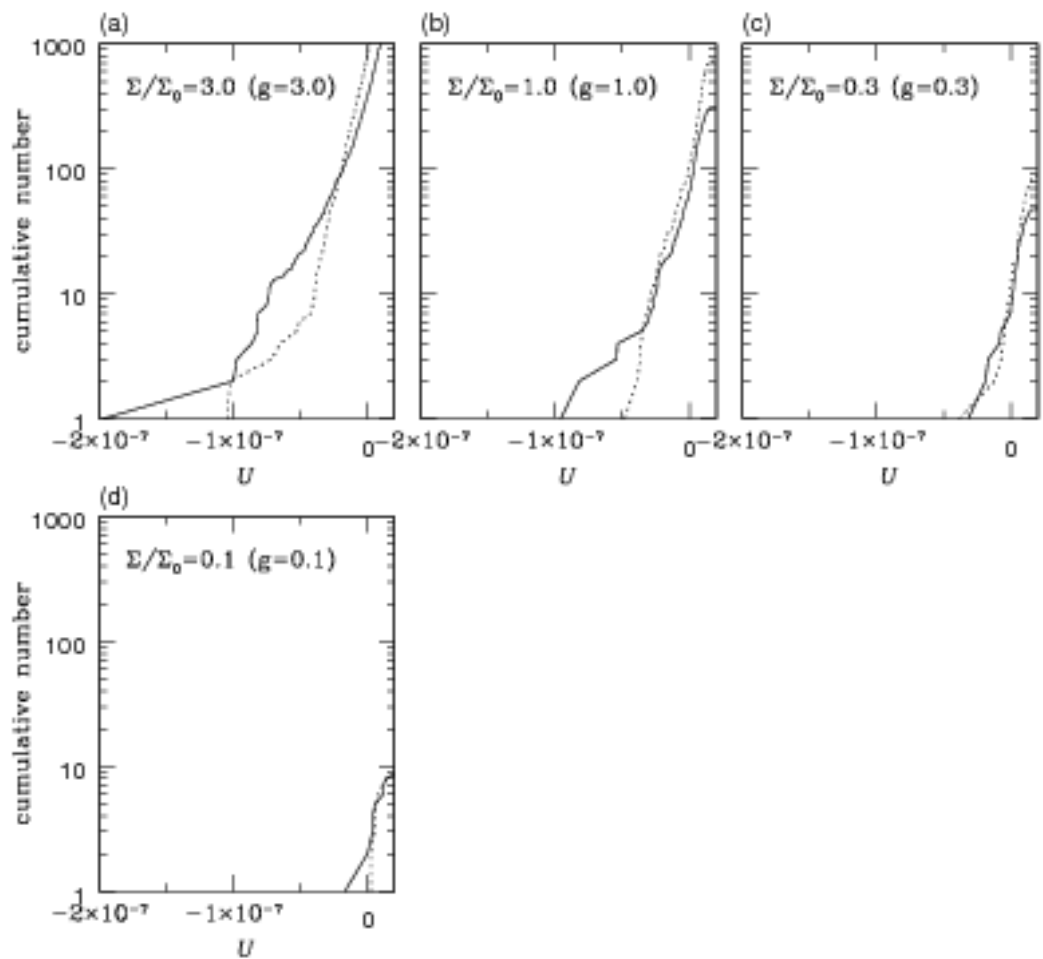

Figure 12: Daisaka.J.K et. al. (2010)

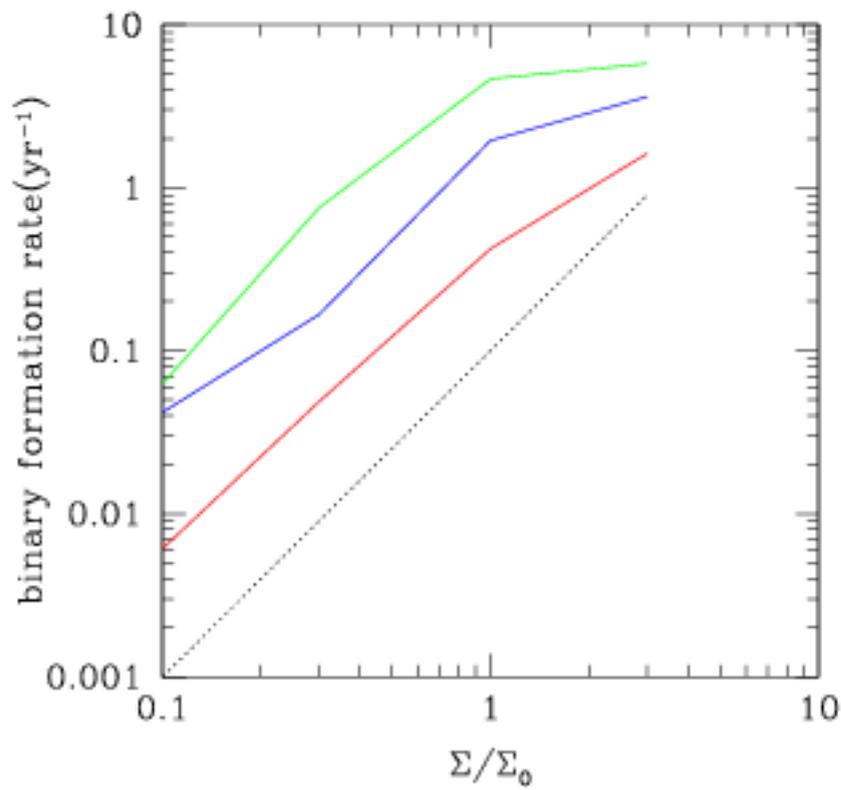

Figure 13: Daisaka.J.K et. al. (2010)

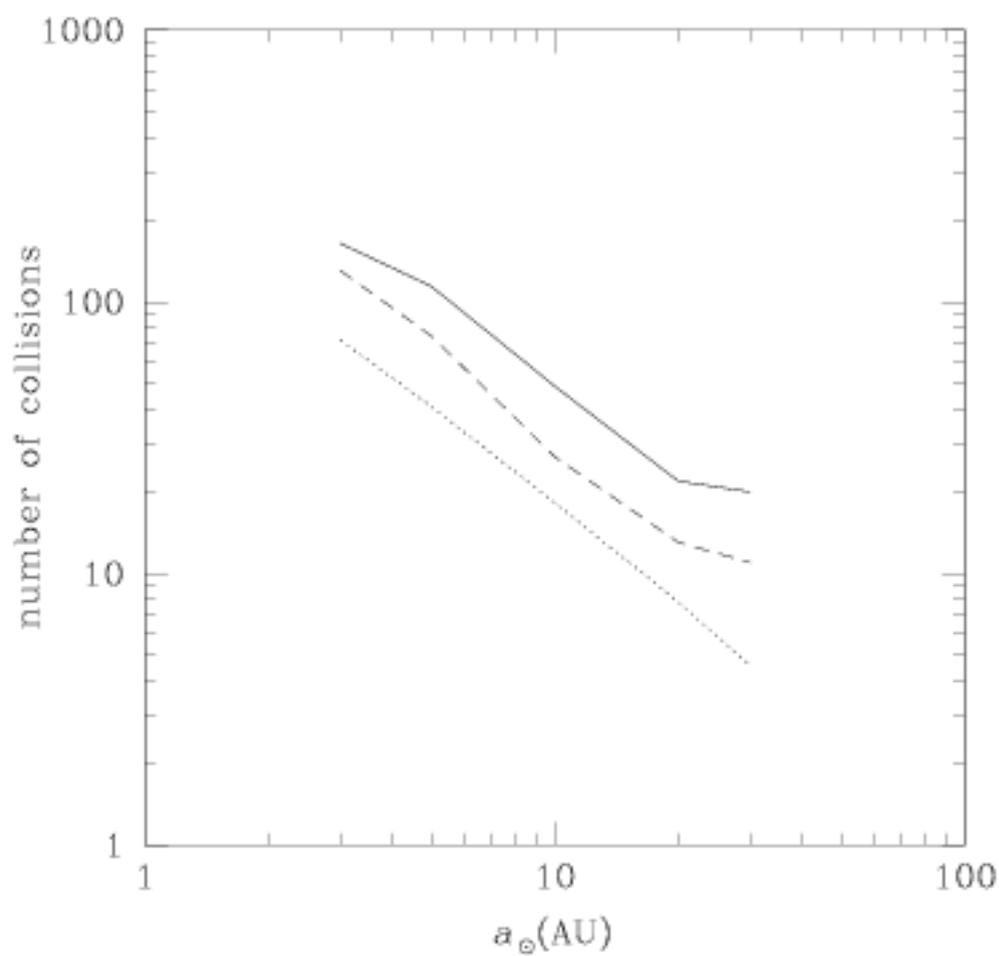

Figure 14: Daisaka.J.K et. al. (2010)

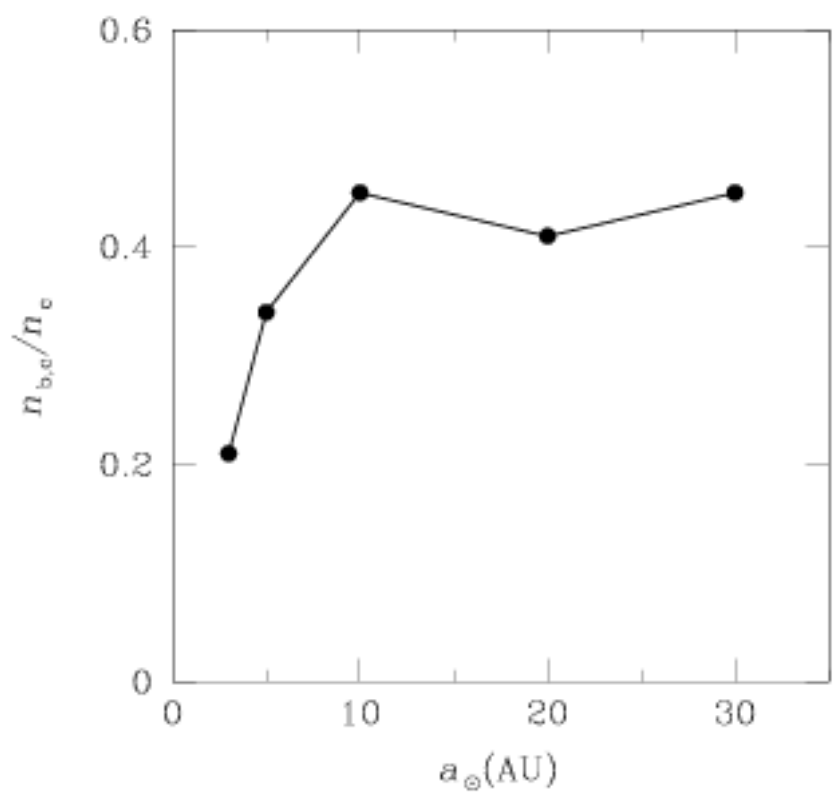

Figure 15: Daisaka.J.K et. al. (2010)

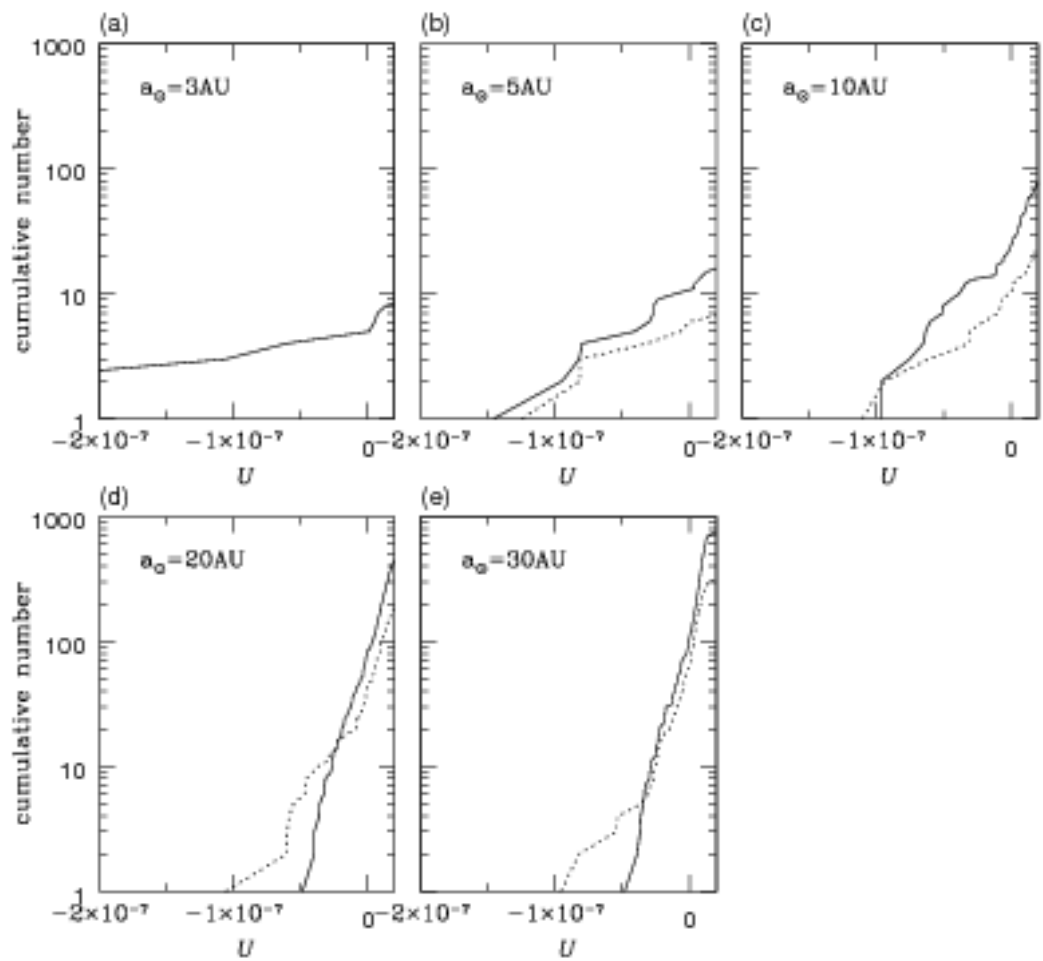

Figure 16: Daisaka.J.K et. al. (2010)